\documentclass[amsmath,amssymb,preprint,superscriptaddress]{revtex4-1}
\usepackage{graphicx}
\begin{document}

\title{Plasma optical modulators for intense lasers}

\author{Lu-Le~Yu}
\email{LuleYu@sjtu.edu.cn}
\affiliation{Key Laboratory for Laser Plasmas (Ministry of Education), Department of Physics and Astronomy, Shanghai Jiao Tong University, Shanghai 200240, China}
\affiliation{Collaborative Innovation Center of IFSA (CICIFSA), Shanghai Jiao Tong University, Shanghai 200240, China}

\author{Yao~Zhao}
\affiliation{Key Laboratory for Laser Plasmas (Ministry of Education), Department of Physics and Astronomy, Shanghai Jiao Tong University, Shanghai 200240, China}
\affiliation{Collaborative Innovation Center of IFSA (CICIFSA), Shanghai Jiao Tong University, Shanghai 200240, China}

\author{Lie-Jia~Qian}
\affiliation{Key Laboratory for Laser Plasmas (Ministry of Education), Department of Physics and Astronomy, Shanghai Jiao Tong University, Shanghai 200240, China}
\affiliation{Collaborative Innovation Center of IFSA (CICIFSA), Shanghai Jiao Tong University, Shanghai 200240, China}

\author{Min~Chen}
\affiliation{Key Laboratory for Laser Plasmas (Ministry of Education), Department of Physics and Astronomy, Shanghai Jiao Tong University, Shanghai 200240, China}
\affiliation{Collaborative Innovation Center of IFSA (CICIFSA), Shanghai Jiao Tong University, Shanghai 200240, China}

\author{Su-Ming~Weng}
\affiliation{Key Laboratory for Laser Plasmas (Ministry of Education), Department of Physics and Astronomy, Shanghai Jiao Tong University, Shanghai 200240, China}
\affiliation{Collaborative Innovation Center of IFSA (CICIFSA), Shanghai Jiao Tong University, Shanghai 200240, China}

\author{Zheng-Ming~Sheng}
\email{zmsheng@sjtu.edu.cn or zhengming.sheng@strath.ac.uk}
\affiliation{Key Laboratory for Laser Plasmas (Ministry of Education), Department of Physics and Astronomy, Shanghai Jiao
Tong University, Shanghai 200240, China}
\affiliation{Collaborative Innovation Center of IFSA (CICIFSA), Shanghai Jiao Tong University, Shanghai 200240, China}
\affiliation{SUPA, Department of Physics, University of Strathclyde, Glasgow G4 0NG, UK}

\author{D. A.~Jaroszynski}
\affiliation{SUPA, Department of Physics, University of Strathclyde, Glasgow G4 0NG, UK}

\author{W. B.~Mori}
\affiliation{University of California, Los Angeles, California 90095, USA}

\author{Jie~Zhang}
\affiliation{Key Laboratory for Laser Plasmas (Ministry of Education), Department of Physics and Astronomy, Shanghai Jiao Tong University, Shanghai 200240, China}
\affiliation{Collaborative Innovation Center of IFSA (CICIFSA), Shanghai Jiao Tong University, Shanghai 200240, China}

\date{\today}

\begin{abstract}
Optical modulators can be made nowadays with high modulation
speed, broad bandwidth, while being compact, owing to the recent
advance in material science and microfabrication technology.
However, these optical modulators usually work for low intensity
light beams. Here, we present an ultrafast, plasma-based optical
modulator, which can directly modulate high power lasers with
intensity up to $10^{16}$ W/cm$^2$ level to produce an extremely
broad spectrum with a fractional bandwidth over 100\%, extending
to the mid-infrared regime in the low-frequency side. This concept
relies on two co-propagating laser beams in a sub-mm-scale
underdense plasma, where a drive laser pulse first excites an
electron plasma wave in its wake while a following carrier laser
beam is modulated by the plasma wave. The laser and plasma
parameters suitable for the modulator to work are presented. Such
optical modulators may enable new applications in the high field
physics.
\end{abstract}

\maketitle

Optical modulators are key components for manipulating optical
signals, which are widely used in scientific  and industrial
applications. For example, high-speed compact electro-optic
modulators (EOMs) are essential for data communications
\cite{Liu15,Reed10,Liu11,Phare15}. EOMs can alter the fundamental
characteristics (i.e., amplitude, frequency, phase and
polarization) of a light beam in a controllable fashion, by making
use of electro-optic effects to change the refractive index of a
material when an external radio-frequency (RF) electric field
driver is applied. Thanks to the rapid development of the field of
RF photonics \cite{Seeds06,Capmany07} together with advanced
material and microfabrication technologies \cite{ Liu11,Phare15,
Melikyan14}, the modulation speed of EOMs has dramatically
increased from megahertz to 100 gigahertz
\cite{Macario12,Chen97,Huang12} over the past decade.
However, it is still very challenging to successfully achieve terahertz (THz) speed for EOMs using current technologies \cite{Liu15}. On the other hand, the rather low optical damage threshold \cite{Bryan84,Furukawa92} and low bulk laser damage threshold of EOMs \cite{Furukawa91} severely limit their applications in the high-intensity regime. For example, a state-of-the-art commercially available magnesium oxide doped LiNbO$_3$ modulator can only handle input light power of $10^2$ mW level and corresponding light intensity at $\sim10^{2}$ W/cm$^2$ \cite{power}.

Currently commercial high power laser systems can deliver peak
powers up to petawatts, which can  be focused to realize laser
intensities from $10^{15}$ W/cm$^2$ to $10^{21}$ W/cm$^2$. The
interaction of such high intensity laser beams with matter is not
only of fundamental interest, but also show prospects of various
applications, such as high harmonic generation \cite{Paul2003},
THz radiation generation \cite{Thomson07,ChenY2008,Cho2015},
plasma-based particle accelerators and light sources
\cite{Esarey09}, laser fusions \cite{Glenzer10}, and laboratory
astrophysics \cite{zhong10}, etc. With such high power lasers, it
has been reported that plasma-based devices have unique advantages
in manipulating intense lasers because they have no damage
threshold. Typical plasma-based optical devices include plasma
channels for the guided propagation of laser pulses over many
Rayleigh lengths \cite{Leemans06b,Leemans14}, plasma mirrors to
improve the temporal contrast of the intense lasers
\cite{Thaury07,Andreev09}, plasma gratings to compress intense
laser pulses\cite{WuHC2005}, plasma lens to focus intense lasers
\cite{WangHY2011}, plasma Raman amplifiers to boost the laser
power to the multipetawatt regime or higher
\cite{Trines11,Mourou06,Yang2015}, and plasma polarization
switching for modulating THz electromagnetic waves \cite{WenH2010}.

\begin{figure}[h]
\begin{center}
\includegraphics[width=3.2in]{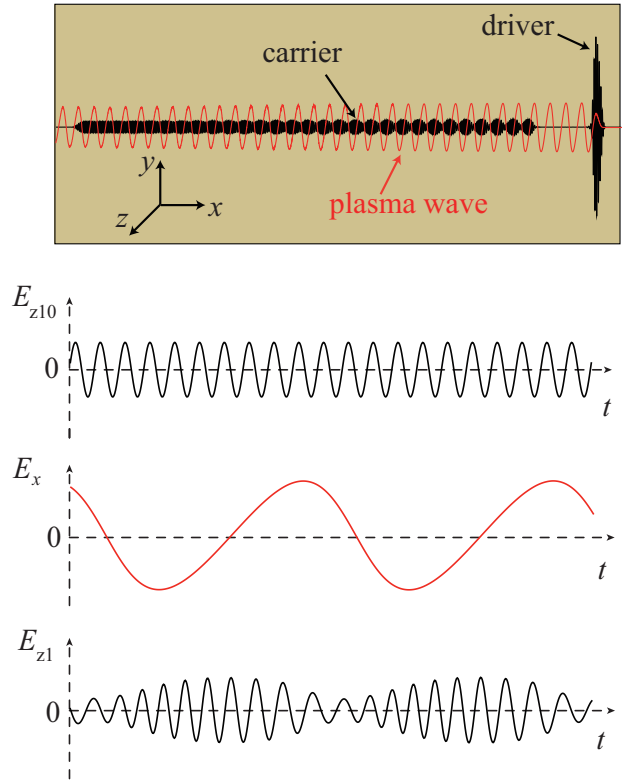}
\end{center}
\caption[example] {\textbf{Schematic of a plasma optical modulator.} An $s$-polarized laser driver propagates (along the $x$ direction) in the plasma and excites an electron plasma wave in its wake. A carrier laser co-propagates behind the driver laser, with a parallel polarization. The amplitude and frequency of the carrier is simultaneously modulated by the plasma wave. Here $E_{z10}$, $E_x$ and $E_{z1}$ are the close-up of the electric fields of the unmodulated carrier, the plasma wave, and the modulated carrier, respectively. The amplitude of $E_x$ has been magnified to be seen clearly.}
\label{Fig1}
\end{figure}

In this article, we show a novel ultrafast all-optical plasma-based
modulator that can efficiently modulate the spectrum of intense
laser pulses to extreme broad bandwidth. The concept is
illustrated in Fig.~\ref{Fig1}: a linearly-polarized femtosecond
intense laser propagates in a sub-mm-scale underdense plasma. The
ponderomotive force of the laser pulse drives the plasma electrons
out of its path. Because the plasma ions are much heavier (by a
fact of at least 1836), they barely move and remain unshielded.
The resultant pattern of alternating positive and negative
charge-separation fields behind the laser driver is a plasma wave,
which has been well-described theoretically \cite{Esarey09} and
measured experimentally \cite{Matlis06}. The wave oscillates at
the plasma frequency $\omega_p$, where
$\omega_p=(n_0e^2/m_e\epsilon_0)^{1/2}$, with $n_0$ the ambient
electron plasma density, $\epsilon_0$ the permittivity of free
space, and $m_e$ and $e$ the electron rest mass and charge,
respectively. For the purpose of optical modulation, the plasma
wave is driven at a moderate amplitude. A picosecond carrier laser
pulse (with arbitrary polarization direction) co-propagates behind
the drive laser with a delay of several plasma wavelengths. The
amplitude and frequency of the carrier are simultaneously
modulated by the plasma wave during its propagation, generating a
number of significant frequency sidebands spaced by the plasma
frequency in the frequency domain. The modulation speed
$f_p$ is determined by the plasma frequency $\omega_p$,
which can be estimated as $f_p ({\rm
Hz})=\omega_p/2\pi\approx8980\sqrt{n_0({\rm cm}^{-3})}$, e.g.,
$f_p=28$ THz for $n_0=10^{19}$ cm$^{-3}$, which is several orders
of magnitude faster than the speed of an EOM. Particle-in-cell
(PIC) simulations show that such plasma modulators can sustain a
carrier intensity up to $10^{16}$ W/cm$^2$, which is several
orders of magnitude higher than what conventional EOMs can handle.

Because of the ultrafast modulation speed and ultrahigh damage
threshold, the plasma optical modulator opens a way to efficiently
modulate laser pulses in the high-intensity regime. Such highly
modulated intense laser pulses may bring a few new physics and
applications associated with intense laser-matter interactions.
For example, it may be used to produce strong THz radiation via
optical rectification as the laser pulses have bandwidth in the
THz range \cite{Rice1994}. Another possibility is to produce
ultra-bright X-ray sources via laser interaction with atoms
\cite{Chang1997}, since the modulated spectrum by our plasma
modulator has been well extended to the mid-infrared regime
\cite{Popmin2012}. The highly modulated laser pulse exhibit
ultrabroad bandwidth, which can suppress the growth rate of the
stimulated Raman scattering (SRS) instability, highly important
for laser fusion \cite{Thomson74,Zhao15}.

\section*{Results}

\begin{figure}[h]
\begin{center}
\includegraphics[width=2.8in]{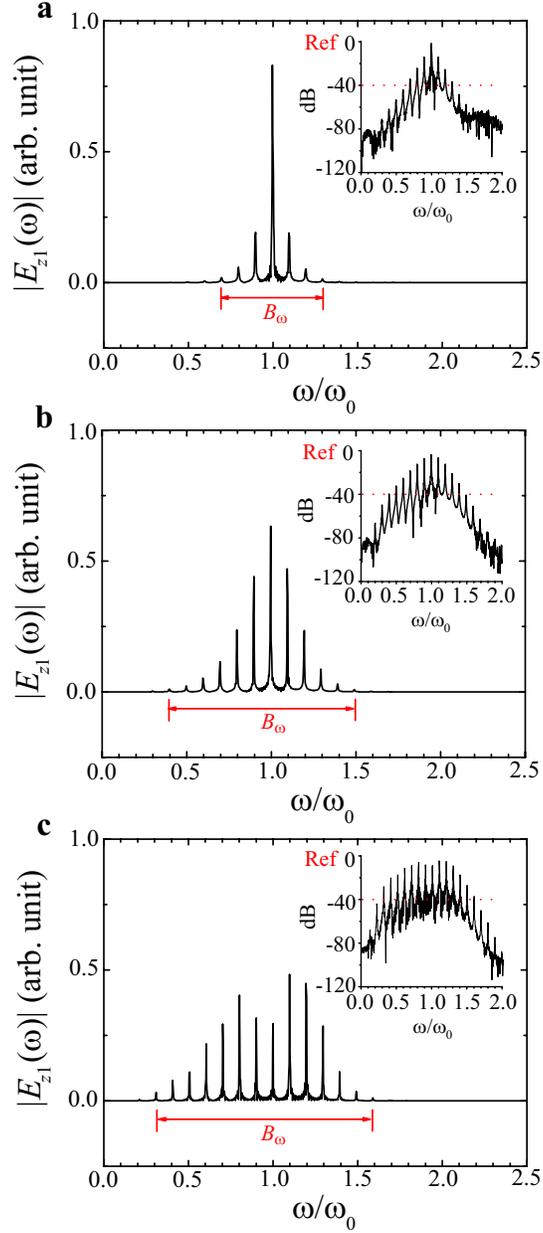}
\end{center}
\caption[example] {\textbf{Evolution of the frequency spectrum of the carrier laser.} The front of the carrier enters the plasma at $t=50T_L$. \textbf{a}, \textbf{b}, and \textbf{c} illustrate the spectra of the carrier at propagation time of $t=250T_L$, $t=400T_L$, and $t=750T_L$ (when the carrier completely passes through the plasma), respectively. $T_L=2\pi/\omega_0$ is the laser cycle in vacuum. The insets in each plot show the spectra using logarithmic coordinates, with 0 dB the reference (i.e., the unmodulated carrier) amplitude. See the laser-plasma simulation parameters in the Methods.}
\label{Fig2}
\end{figure}

Figure~\ref{Fig2} shows an example case to demonstrate the
essential features of the modulation obtained  from
one-dimensional (1D) PIC simulations. For simplicity, the vacuum
wavelengths of the two laser pulses are both 1 $\mu$m in the
simulations. In practice, an 800 nm Ti:sapphire femtosecond laser
pulse can be used to excite the plasma waves, which does not lead
to obvious changes of the results presented in the following. The
normalized field amplitudes of the driver and the unmodulated
carrier are $a_{00}=0.8$ and $a_{10}=0.05$, respectively, where
$a_{i0}=E_{zi0}/E_n~(i=0,1)$, and $E_n=m_ec\omega_0/e$, with $c$
and $\omega_0$ the light speed and angular frequency in vacuum,
respectively. For linear polarization, $I_{i0}({\rm W/cm}^{2})=
1.37\times10^{18}a_{i0}^2/[\lambda_{0}(\mu {\rm m})]^2$, with
$I_{i0}$ the peak laser intensity and $\lambda_{0}=2\pi
c/\omega_0$ the laser wavelength in vacuum. Thus, $a_{00}$ and
$a_{10}$ correspond to the laser intensities of
$8.77\times10^{17}$ W/cm$^2$ and $3.43\times10^{15}$ W/cm$^2$,
respectively, for 1 $\mu$m laser wavelength. Detailed parameters
are given in the Methods. The modulated pulse can be
well-described using the analytical model presented in the
Methods. It can be expressed as
$a_1(t)=a_{10}[1+m\cos(\omega_pt)]\cos[\omega_0t+\beta\sin(\omega_pt)]$,
where $m$ and $\beta$ are the amplitude modulation index and the
frequency modulation index, respectively. The mixed amplitude and
frequency modulation of a sinusoidal carrier by a simple
sinusoidal plasma wave yields a mass of sidebands including both
Stokes and anti-Stokes components given by $\omega_n=\omega_0\pm
n\omega_p$ (with $n$ a nonzero integer and $\omega_p$ as a
frequency interval). Note that in the quasi-linear
($a_{i0}^2\lesssim1$) regime, i.e., where the
relativistic-electron-mass increase associated with the motion of
the plasma electrons can be neglected, the frequency $\omega_p$
can be calculated as $\omega_p=(n_0/n_c)^{1/2}\omega_0$, with $n_c
({\rm cm}^{-3})=1.1\times10^{21}/[\lambda_{0}(\mu {\rm m})]^2$ the
critical plasma density for the corresponding incident laser
wavelength $\lambda_0$. The spectral bandwidth is defined as
$B_{\omega}=2(\beta+1)\omega_p$ \cite{sideband}, where $\beta$
depends upon the amplitude of driver pulse and the plasma density,
in addition to the plasma length.

For high fidelity, we only count the significant sidebands with
the amplitudes larger than 1 percent (-40 dB) of the amplitude of
the unmodulated carrier \cite{sideband}. Therefore, the spectral
bandwidth of the modulated carrier can be calculated by estimating
the number of significant sidebands. As shown in Fig.~\ref{Fig2},
the higher-order sidebands gradually grow with the laser-plasma
interaction time. When the carrier beam completely passes through
the plasma, the maximum significant sidebands for the anti-Stokes
and Stokes components are $\omega_{+6}$ and $\omega_{-7}$,
respectively, giving a bandwidth of
$B_{\omega}=13\omega_p=1.3\omega_0$, accounting for the fact that
$\omega_p=0.1\omega_0$ for $n_0/n_c=0.01$. It is also noted that
the sideband spectrum of a mixed modulation is asymmetrical due to
the superposition of the sideband components of both amplitude and
frequency modulations. The simulation results are in good
agreement with the prediction of the analytical model given in the
Methods. In this example, the amplitude and frequency modulation
indices can be estimated \cite{sideband} as $m=(a_{10,{\rm
max}}-a_{10,{\rm min}})/(a_{10,{\rm max}}+a_{10,{\rm min}})=0.42$,
and $\beta=B_{\omega}/(2\omega_p)-1=5.5$, respectively. Note that
$\beta\gg 1$ corresponds to broadband modulation. The energy
transmission rate of the carrier through plasma is about 94.3\% in
this example.

\begin{figure}[h]
\begin{center}
\includegraphics[width=6.5in]{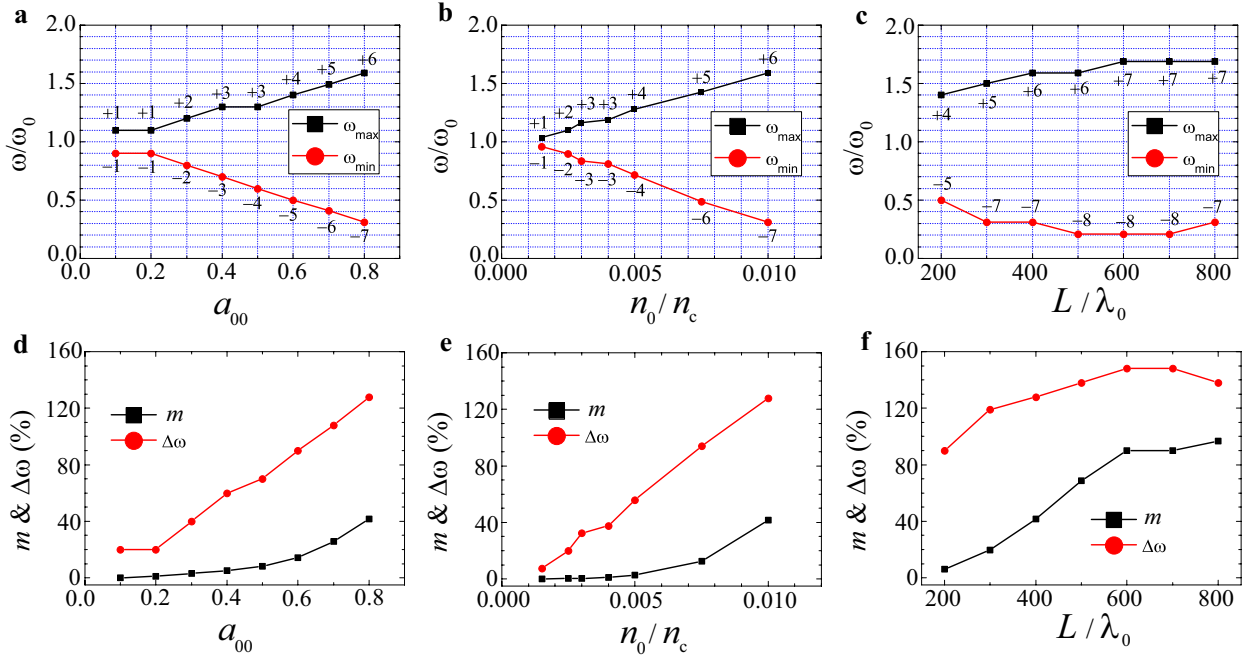}
\end{center}
\caption[example] {\textbf{Performance metrics of the plasma optical modulator.} \textbf{a}-\textbf{c}, The -40 dB sidebands versus (\textbf{a}) the driver amplitude, (\textbf{b}) the plasma density, and (\textbf{c}) the plasma length. \textbf{d}-\textbf{f}, The -40 dB fractional bandwidth $\Delta\omega$ and the amplitude modulation index $m$ versus (\textbf{d}) the driver amplitude, (\textbf{e}) the plasma density, and (\textbf{f}) the plasma length. The other laser-plasma parameters are the same as in Fig.~\ref{Fig2}.}
\label{Fig3}
\end{figure}

We find the modulation is effective for a wide range of
laser-plasma parameters. Figure~\ref{Fig3} shows the -40 dB
cutoff sidebands, the corresponding fractional bandwidth
($\Delta\omega=B_{\omega}/\omega_0$), and the amplitude modulation
index $m$, as a function of the driver intensity, plasma density
and plasma length. When the driver amplitude is relatively small
(e.g., $a_{00}=0.1$), the modulation is quite weak so that the
spectrum only consists of the 1st-order sidebands. By increasing
the driver amplitude, the field strength of the plasma wave is
enhanced, and subsequently the modulation indices become larger,
leading to higher-order sidebands and a wider bandwidth. The
similar scaling law exists when increasing the plasma density.
Therefore, by properly increasing the driving laser intensity and
plasma density, one can extend the spectrum of the modulated
carrier to the mid-infrared regime in the low frequency side (or
the Stokes waves). We note that the growth of the bandwidth is
relatively insensitive to the increase of the plasma length after
certain distance, which implies a saturation of modulation. As
shown in Fig.~\ref{Fig3}d-\ref{Fig3}f, the amplitude modulation
index $m$ gradually grows with the increase of the driver
intensity or the plasma density. When increasing the plasma
length, $m$ first grows and then saturates at the 100\% level,
which indicates that the carrier breaks up into a train of short
pulses, and each of these short pulses has a width on the order of
the plasma wavelength. According to Fig.~\ref{Fig3}d-\ref{Fig3}f,
$\Delta\omega$ and $\beta$ have similar dependence on the driver
intensity, the plasma density, and the plasma length as $m$.

\begin{figure}[h]
\begin{center}
\includegraphics[width=2.8in]{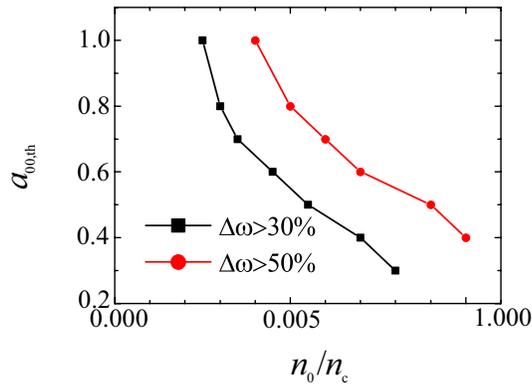}
\end{center}\caption[exmple] {\textbf{Threshold driver amplitude versus the plasma density for generating broad-bandwidth carriers with fractional bandwidth over 30\% and 50\%.} The laser-plasma parameters are the same as in Fig.~\ref{Fig2}.}
\label{Fig4}
\end{figure}

For certain applications, it is important to know the parameter
range for  broad bandwidth generation. Figure~\ref{Fig4}
illustrates the parameter range for generating ultrabroad
bandwidths (for example, $\Delta\omega\geqslant30\%$), which
presents a series of simulations where the threshold for the
driver amplitude $a_{00,\mathrm{th}}$ is scanned for a given
plasma density $n_0$. In general, a broader bandwidth can be
achieved at a higher plasma density even if a lower driver
intensity is adopted.

\section*{Discussion}

So far we have discussed the spectrum development of the carrier
laser pulse as a function of the drive laser amplitude, plasma
density, and the plasma length. One question still to be answered
is the maximum carrier laser intensity allowed in the plasma
modulator. A previous study has shown that the resultant pulse
train can amplify the field strength of the plasma wake to a wave
breaking level if the initial amplitude of carrier laser intensity
is high enough \cite{Sheng02}. We also find the remarkably
enhanced plasma waves when the intensity of the pulse train is on
the same order of the drive intensity (e.g., $10^{17}$ W/cm$^2$
level). This can result in severe distortion of the plasma wave
and considerable energy loss of the carrier laser to plasma wave
excitation as well as electron trapping and acceleration (see
Supplementary information). As a consequence, the frequency
modulation of the carrier laser is suppressed. Therefore, the
maximum intensity of the carrier laser should be well below
$10^{17}$ W/cm$^2$ (e.g., at $10^{16}$ W/cm$^2$ level) in order to
realize an excellent performance of the plasma optical modulator.

The maximum allowed pulse duration of the carrier laser for
effective modulation may be interesting for some particular
applications. This depends upon the life time of the electron
plasma waves, which is determined by the collisional damping,
Landau damping, and phase mixing \cite{Gupta99,HXu06}. Typically
the initial electron plasma temperature $T_e$ is over 10 eV and
the effective $T_e$ is over 100 eV when considering the electron
quiver motion in the carrier laser with intensity about $10^{16}$
W/cm$^2$, which leads to a time scale of over 10 ps for the
collisional damping under the plasma electron density of
$\sim10^{19}{\rm cm}^{-3}$. The Landau damping time is much longer than
the collisional damping time in the present case when the plasma wave
is driven at moderate amplitudes. The phase mixing due to the ion
motion is the key responsibility for the plasma wave decay since
it occurs on a much shorter time scale of
$t_{\mathrm{mix}}\sim\omega_p^{-1}[(A^2/24)(m_e/m_i)]^{-1/3}\sim1.47$
ps \cite{Gupta99} when a high Z gas such as argon is used
for the plasma wave generation, where $A=E_{\rm max}/E_p\sim a^2_{00}/2$ is the
normalized amplitude of the plasma wave, with $E_p=cm_e\omega_p/e$
and $m_i$ the ion mass. PIC simulations show that the
plasma wave starts to decay around 1.65 ps due to the phase
mixing, which is in good agreement with the analytical model. The
maximum pulse duration for the effective modulation is around 3 ps
for the laser-plasma parameters under consideration (see
Supplementary information).

\begin{figure}[h]
\begin{center}
\includegraphics[width=5.0in]{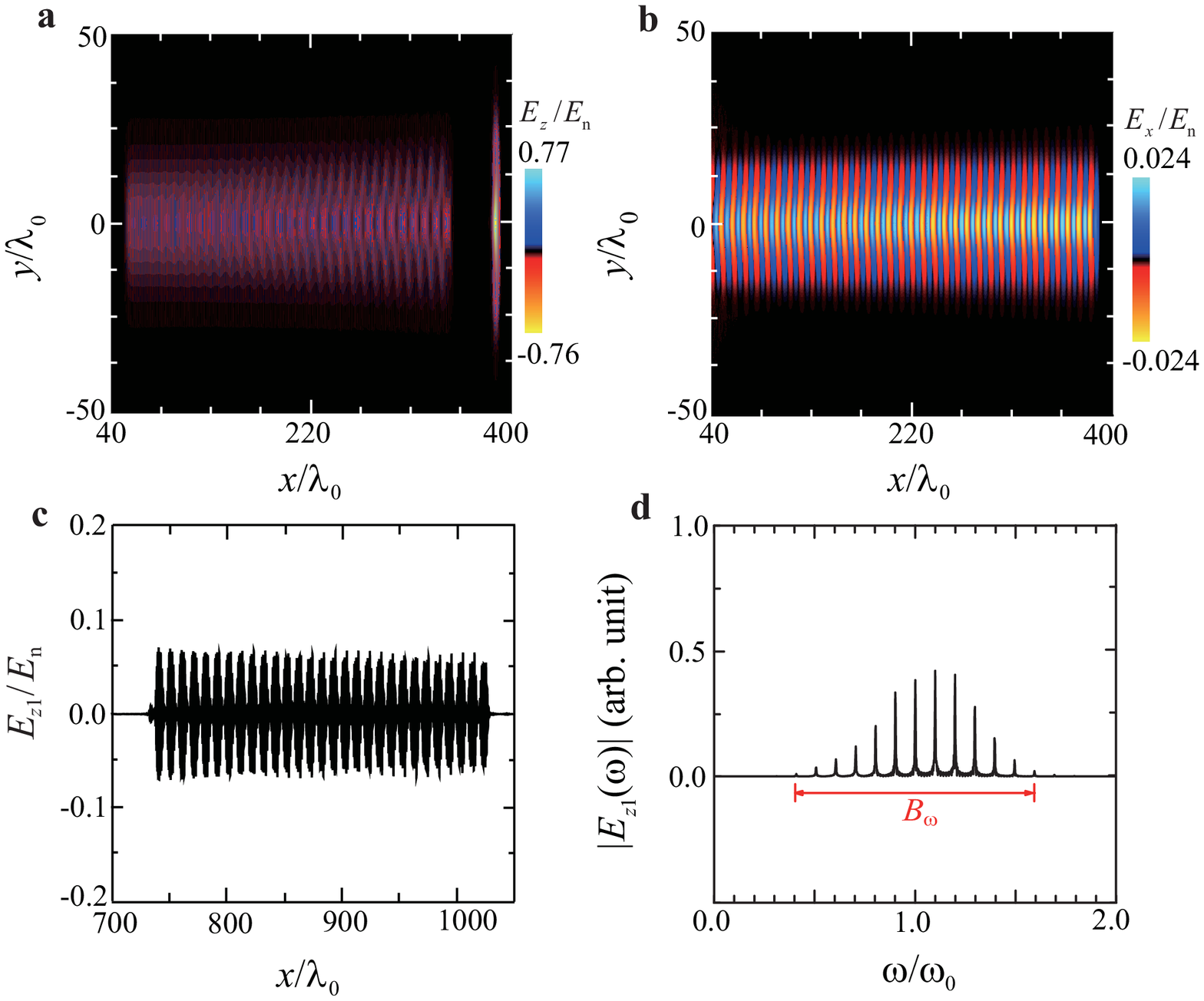}
\end{center}
\caption[example] {\textbf{Two-dimensional simulation results.} Snapshots of (\textbf{a}) the electric field of the laser pulses, (\textbf{b}) the longitudinal electric field of the plasma wave, at $t=392T_L$, (\textbf{c}) the on-axis electric field of the carrier, and (\textbf{d}) the corresponding frequency spectrum of the field, when it completely passes through the plasma. Laser-plasma simulation parameters are given in the Methods.}
\label{Fig5}
\end{figure}

Another issue is the spot sizes of the laser pulses. As we have
shown above, the laser pulses need to propagate over a distance
about 1 mm without significant transverse spreading. One needs to
take relatively large laser spot size so that
the corresponding Rayleigh length $\pi r_0^2/\lambda_0$ is long
enough. Meanwhile, self-focusing will occur when the driver power
$P$ exceeds a critical power $P_c$, with $P_c~({\rm
GW})=17.4(\omega_0/\omega_p)^2$. For linear polarization,
$P/P_c=(\omega_pr_0a_{00})^2/(32c^2)$\cite{Esarey09}. To avoid
strong self-focusing within 1 mm, the laser spot size also cannot
be too large. Two-dimensional (2D) simulations show that the
optimal modulation can be achieved for $1\lesssim P/P_c\lesssim2$.
An example of 2D simulation is given in Fig.~\ref{Fig5}. Detailed
parameters are given in the Methods. In this example, the driver
power is $P/P_c=1.18$ and weak self-focusing occurs during the
propagation. As shown in Fig.~\ref{Fig5}a, the maximum amplitude of the driver increases
by 10\% (from $a_{00}=0.7$ to 0.77) at a propagation
distance of $392\lambda_0$. The excited plasma wave
retains the quasi-1D structure, and keeps quite stable during
propagation, which is advantageous for the modulation process. It
is noted that the driver spot leads to transverse inhomogeneity of
the plasma wave, resulting in transverse inhomogeneity of the
modulation. By reducing the driver intensity, the corresponding
spot size can be increased, and hence, the transverse uniformity
of the modulation can be improved.
A test experiment can be done with a carrier laser pulse [e.g.,
Nd:YVO$_4$, 1 $\mu$m, $\sim$ 15 mJ, $\sim$ 1 ps in the full width at
half maximum (FWHM) of the intensity] delayed
with respect to a drive laser pulse [e.g., Ti:sapphire, 0.8
$\mu$m, $\sim$ 70 mJ, $\sim$ 20 fs (FWHM)], co-propagating in a 1-mm-long helium gas with
a density of $\sim 10^{19}$ cm$^{-3}$. The time delay between the
two laser pulses, which is relatively flexible, can be controlled in a timescale of hundreds of
femtoseconds. Therefore, the plasma optical modulator can be
tested experimentally without significant technical difficulties.

In summary, we have illustrated a novel application of the plasma
wave as a unique optical modulator for intense lasers. It relies
on two co-propagating laser pulses in a short underdense plasma: A
driver with a typical intensity $10^{17}$ W/cm$^2$, which propagates
in the plasma and excites a plasma wake, and a carrier, which
propagates behind the driver by several plasma wavelengths. Both
the amplitude and frequency of the carrier are modulated by the
plasma wave, leading to an ultrabroad bandwidth in its spectrum
which extends to the mid-infrared range. The modulation speed is
in the THz regime. Compared with the low damage threshold of the
conventional EOMs, the plasma modulator allows the carrier
intensity as high as up to $10^{16}$ W/cm$^2$. In addition, the
plasma modulator offers excellent performance control by changing the driver intensity,
the plasma density, and the plasma length.
The required experimental conditions for such plasma modulators are
within current technical capabilities.

\section*{Methods}

\subsection{Mathematical model for mixed-modulation.}

Physically, the modulation of the carrier laser pulse by an
electron plasma wave is similar to that found for an intense laser
propagation in plasma via stimulated Raman forward scattering
(coupled with the self-modulation instability)
\cite{mori1997,tzeng1999}. The latter leads to a spectrum of
Stokes and anti-Stokes waves when the laser pulse has a duration
longer than a plasma wavelength. The evolution of the amount of
Stokes/anti-Stokes modes can be described by photon acceleration
and deceleration \cite{wilks1989,esarey1990,sheng1993}. The
dependence of the spectral modulation on the plasma wave
amplitude, plasma density and interaction time discussed in the
above section also qualitatively agrees with the previous
theories. The difference between the spectral modulation described
in the previous theories and here is that our plasma optical modulator
enables the spectral modulation of the carrier laser to be
well-controlled and to be developed much more efficiently. The
frequency modulation due to stimulated Raman forward scattering
and self-modulation instabilities can be ignored in our case.

The carrier pulse is modulated in the amplitude and frequency by the electron plasma wave, i.e., a mixed modulation.
Its temporal structure can be written as\cite{Ozimek87}
\begin{eqnarray}\label{eq1}
a_1(t)=&&a_{10}[1+m\cos(\omega_pt)]\nonumber\\
&&\times\cos[\omega_0t+\beta\sin(\omega_pt)]\nonumber,
\end{eqnarray}
assuming that the excited plasma wave is a simple sinusoidal
oscillation, with the normalized axial electric field
$E_x/E_p=(E_{\rm max}/E_p)\cos(\omega_pt)$\cite{Esarey09}.
Here $a_{10}$ is the normalized amplitude of the unmodulated carrier. For a linearly-polarized
sinusoidal driver with optimal length for plasma wave excitation
(i.e., the pulse length approximate to the plasma wavelength),
$E_{\rm max}/E_p\sim a^2_{00}/2$, yielding the amplitude and
frequency modulation indices $m\varpropto a^2_{00}/(2a_{10})$ and
$\beta\varpropto (a^2_{00}/2)(\omega_0/\omega_p)$, respectively.
These two parameters also depend upon the interaction time or the
plasma length as shown in the simulation results given in Fig. 3.
Using simple trigonometrical transformations and a lemma of Bessel
function\cite{Kreh12} and $J_{-n}(\beta)=(-1)^nJ_n(\beta)$,
$a_1(t)$ can be written as
\begin{eqnarray}\label{eq2}
a_1(t)&&=a_{10}\sum_{n=-\infty}^{\infty}J_{n}(\beta)\cos(\omega_0t+n\omega_pt)\nonumber\\
&&+\frac{1}{2}a_{10}m\sum_{n=-\infty}^{\infty}J_{n}(\beta)\cos(\omega_0t+\omega_pt+n\omega_pt)\nonumber\\
&&+\frac{1}{2}a_{10}m\sum_{n=-\infty}^{\infty}J_{n}(\beta)\cos(\omega_0t-\omega_pt+n\omega_pt)\nonumber.
\end{eqnarray}
It is obvious that the spectrum of $a_1(t)$ primarily consists of
three components: the central frequency $\omega_0$ that
corresponds to the unmodulated carrier, and the two first-order sidebands
$\omega_{\pm1}=\omega_0\pm\omega_p$ resulting from the modulation
process. The amplitudes of the frequency components can be
characterized by the expansion in a series of $n$th-order Bessel
function $J_n$. By taking Fourier transformation of $a_1(t)$, and
considering the $k$th-order frequency component with $k$ a
positive integer, we can get the amplitudes of the upper sideband
($\omega_{+k}=\omega_0+k\omega_p$) and the lower sideband
($\omega_{-k}=\omega_0-k\omega_p$) in the spectrum as follows:
\begin{eqnarray}\label{eq3}
|a_1(\omega_{+k})|=&&(\sqrt{2\pi}/2)a_{10}|[J_{k}(\beta)+(m/2)J_{k-1}(\beta)\nonumber\\
&&+(m/2)J_{k+1}(\beta)]|\nonumber,\\
|a_1(\omega_{-k})|=&&(\sqrt{2\pi}/2)a_{10}|[J_{k}(\beta)-(m/2)J_{k-1}(\beta)\nonumber\\
&&-(m/2)J_{k+1}(\beta)]|\nonumber,
\end{eqnarray}
for $\omega_{\pm k}>0$.
From the above equations, it is obvious that the amplitude of the lower sideband is not equal to the amplitude of the corresponding upper sideband, resulting in an asymmetrical sideband spectrum.
For a weak frequency modulation ($0<\beta\ll1$), the modulation index is so small that the spectrum essentially consists of $\omega_0$ and only one set of sidebands $\omega_{\pm1}$, with the amplitudes of  $|a_1(\omega_{+1})|=(\sqrt{2\pi}/4)(m+\beta)a_{10}$,
and $|a_1(\omega_{-1})|=(\sqrt{2\pi}/4)|m-\beta|a_{10}$. For a large modulation index ($\beta>1)$, there will be a number of significant sidebands spanning over a broad frequency range.

\subsection{PIC Simulations.}
Simulations have been carried out using the code {\sc osiris}
\cite{Fonseca02}. In the 1D simulations (e.g., in
Fig.~\ref{Fig2}), the temporal profile of the drive pulse is
$a_0(t)=a_{00}\sin^2(\pi t/T_0)$, with $0\leq t\leq T_0$ and
$T_0=10T_L$. The carrier pulse, which is delayed by $40\lambda_0$
from the driver, has a duration of $T_1=303T_L$. It has a similar
profile as the driver at its leading and trailing edges, and a
plateau of $283T_L$ in between. The amplitudes of the driver and
the carrier are $a_{00}=0.8$, $a_{10}=0.05$, respectively. The
trapezoid-shaped plasma has a length of $400\lambda_0$ with a
plateau of $380\lambda_0$, located between
$x=10\lambda_0$ and $x=410\lambda_0$. The initial plasma electron
density in the plateau region is set to be $n_0/n_c=0.01$. For
laser-driven plasma waves, typically the initial (photoionized)
electron plasma temperature is set to be 10 eV. The simulation box
size is $800\lambda_0$ with 20 macro-particles per cell. The
resolution of the computational grid is $\Delta x=\lambda_0/40$.
At $t=0$, the front of the driver enters the simulation box. In
the 2D simulation (Fig.~\ref{Fig5}), the amplitude of the driver is $a_{00}=0.7$
and the spot sizes of the driver
and the carrier are $w_0=14\lambda_0$ and $w_1=17\lambda_0$,
respectively. The trapezoid-shaped plasma has a length of
$700\lambda_0$. Other laser-plasma parameters are the same as the
1D simulations. The simulation box size is
$1100\lambda_0\times100\lambda_0$ with 4 macro-particles per cell.
The resolution of the computational grid is $\Delta
x=\lambda_0/32$ and $\Delta y=\lambda_0/20$.

\section*{Acknowledgements}
The authors would like to thank Thomas Sokollik, Yulong Tang, Jun
Zheng, Yanping Chen, and Guoqiang Xie for useful discussions. This
work was supported by the National Basic Research Program of China
under Grant No.~2013CBA01500, 2014CB339801 and 2015CB859700, the
National Natural Science Foundation of China under Grants
No.~11421064, 11405107 and 11475113. Simulations have been carried
out on the PI supercomputer at Shanghai Jiao Tong University and
the Milky Way 2 supercomputer in the National Supercomputer Center
in Guangzhou. We acknowledge the support of the UK EPSRC (grant
no. EP/J018171/1) and a Leverhulme Trust Research Project Grant.

\section*{Author contributions}
L.L.Y. and Z.M.S. designed the overall concept presented in this paper. L.L.Y. carried out all the simulations and the analytical model, and wrote the main manuscript text. Z.M.S., Y.Z., L.J.Q., M.C., S.M.W., D.A.J., W.B.M., and J.Z.
contributed to analyze the results and write the manuscript.
All authors discussed the results and commented on the manuscript.

\section*{Additional information}
Supplementary information is available in the online version of the paper.
Reprints and permissions information is available online at www.nature.com/reprints. Correspondence and
requests for materials should be addressed to L.L.Y. and Z.M.S.

\section*{Competing financial interests}
The authors declare no competing financial interests.

\clearpage

\section*{Figure legends}

\section*{References}
\begin{thebibliography}{47}
\bibitem{Liu15}
Liu, K., Ye, C. R., Khan, S. $\&$ Sorger, V. J. Review and perspective on ultrafast wavelength-size electro-optic modulators. \textit{Laser Photonics Rev.} \textbf{9}, 172-194 (2015).

\bibitem{Reed10}
Reed,G. T., Mashanovich, G., Gardes, F. Y. $\&$ Thomson, D. J. Silicon optical modulators. \textit{Nat. Photon.} \textbf{4}, 518-526 (2010).

\bibitem{Liu11}
Liu, M. \textit{et al}. A graphene-based broadband optical modulator. \textit{Nature} \textbf{474}, 64-67 (2011).

\bibitem{Phare15}
Phare, C. T., Lee, Y.-H. D., Cardenas, J. $\&$ Lipson, M. Graphene electro-optic modulator with
30 GHz bandwidth. \textit{Nat. Photon.} \textbf{9}, 511-515 (2015).

\bibitem{Seeds06}
Seeds, A. J. $\&$ Williams, K. J. Microwave photonics. \textit{J. Lightwave Technol.} \textbf{24}, 4628-4641 (2006).

\bibitem{Capmany07}
Capmany, J. $\&$ Novak, D. Microwave photonics combines two world. \textit{Nat. Photon.} \textbf{1}, 319-330 (2007).

\bibitem{Melikyan14} Melikyan, A. \textit{et al}. High-speed plasmonic phase modulators. \textit{Nat. Photon.} \textbf{8},
229-233 (2014).

\bibitem{Macario12}
Macario, J. \textit{et al}. Full spectrum millimeter-wave modulation. \textit{Opt. Express} \textbf{20}, 23623-23629 (2012).

\bibitem{Chen97}
Chen, D. \textit{et al}. Demonstration of 110 GHz electro-optic polymer modulators. \textit{Appl. Phys. Lett.} \textbf{70}, 3335-3337 (1997).

\bibitem{Huang12}
Huang, H. \textit{et al}. Broadband modulation performance of 100-GHz EO polymer MZMs. \textit{J. Lightwave Technol.} \textbf{30}, 3647-3652 (2012).


\bibitem{Bryan84}
Bryan, D. A., Gerson, R. $\&$ Tomaschke, H. E. Increased optical damage resistance in lithium niobate. \textit{Appl. Phys. Lett.} \textbf{44}, 847-849 (1984).

\bibitem{Furukawa92}
Furukawa, Y., Sato, M., Kitamura, K., Yajima, Y. $\&$ Minakata, M. Optical damage resistance and crystal quality of LiNbO3 single crystals with various [Li]/[Nb] ratios. \textit{J. Appl. Phys.} \textbf{72}, 3250-3254 (1992).


\bibitem{Furukawa91}
Furukawa, Y. \textit{et al}. Investigation of bulk laser damage threshold of lithium niobate single crystals by
Q-switched pulse laser. \textit{J. Appl. Phys.} \textbf{69}, 3372-3374 (1991).


\bibitem{power}
\url{http://www.newport.com/Electro-Optic-Modulator-Selection-Guide/977460/1033/content.aspx?xcid=bing-ppc-0244}


\bibitem{Paul2003} Paul, A. \textit{et al}. Quasi-phase-matched generation of coherent extreme-ultraviolet light.
\textit{Nature} \textbf{421}, 51-54 (2003).

\bibitem{Thomson07} Thomson, M. D., Kre{\ss}, M., L\"{o}ffler, T. $\&$ Roskos, H. G. Broadband THz emission from gas plasmas
induced by femtosecond optical pulses: From fundamentals to applications. \textit{Laser $\&$ Photon. Rev.} \textbf{1}, 349-368 (2007).

\bibitem{ChenY2008} Chen, Y. \textit{et al}. Elliptically polarized Terahertz emission in the forward direction of a femtosecond
laser filament in air. \textit{Appl. Phys. Lett.} \textbf{93}, 231116 (2008).

\bibitem{Cho2015} Cho, M.-H. \textit{et al}. Strong terahertz emission from electromagnetic diffusion near cutoff in
plasma.\textit{New J. Phys.} \textbf{17}, 043045 (2015).


\bibitem{Esarey09}
Esarey, E., Schroeder, C. B. $\&$ Leemans, W. P. Physics of laser-driven plasma-based electron accelerators. \textit{Rev. Mod. Phys.} \textbf{81}, 1229-1285 (2009).

\bibitem{Glenzer10}
Glenzer, S. H. \textit{et al}. Symmetric inertial confinement fusion implosions at ultra-high laser energies. \textit{Science} \textbf{327}, 1228-1231 (2010).

\bibitem{zhong10}
Zhong, J.-Y. \textit{et al}. Modelling loop-top X-ray source and reconnection outflows in solar flares with intense lasers.
\textit{Nat. Phys.} \textbf{6}, 984-987 (2010).


\bibitem{Leemans06b}
Leemans, W. P. \textit{et al}. GeV electron beams from a centimetre-scale accelerator. \textit{Nat. Phys.} \textbf{2}, 696-699 (2006).

\bibitem{Leemans14}
Leemans, W. P. \textit{et al}. Multi-GeV electron beams from capillary-discharge-guided subpetawatt laser pulses in the self-trapping regime. \textit{Phys. Rev. Lett.} \textbf{113}, 245002 (2014).

\bibitem{Thaury07}
Thaury, C. \textit{et al}. Plasma mirrors for ultrahigh-intensity optics. \textit{Nat. Phys.} \textbf{3}, 424-429 (2007).

\bibitem{Andreev09}
Andreev, A. A. \textit{et al}. Optimal ion acceleration from ultrathin foils irradiated by a profiled laser
pulse of relativistic intensity. \textit{Phys. Plasmas} \textbf{16}, 013103 (2009).

\bibitem{WuHC2005} Wu, H.-C., Sheng, Z.-M., Zhang, Q.-J., Cang, Y. $\&$  Zhang, J. Manipulating ultrashort intense laser pulses
by plasma Bragg gratings. \textit{Phys. Plasmas} \textbf{12}, 113103 (2005).


\bibitem{WangHY2011} Wang, H.-Y. \textit{et al}. Laser shaping of a relativistic intense, short Gaussian pulse by a plasma lens.
\textit{Phys. Rev. Lett.} \textbf{107}, 265002 (2011).

\bibitem{Trines11}
Trines, R. M. G. M. \textit{et al}. Simulations of efficient Raman amplification into the multipetawatt regime.
\textit{Nat. Phys.} \textbf{7}, 87-92 (2011).

\bibitem{Mourou06}
Mourou, G. A., Tajima, T. $\&$ Bulanov, S. V. Optics in the relativistic regime. \textit{Rev. Mod. Phys.} \textbf{78}, 309-371 (2006).

\bibitem{Yang2015} Yang, X. \textit{et al}. Chirped pulse Raman amplification in warm plasma: towards controlling saturation.
\textit{Sci. Rep.} \textbf{5}, 13333 (2015).


\bibitem{WenH2010} Wen, H., Daranciang, D. $\&$ Lindenberg, A. M. High-speed all-optical terahertz polarization switching by a
transient plasma phase modulator. \textit{Appl. Phys. Lett.} \textbf{96}, 161103 (2010).

\bibitem{Matlis06} Matlis, N. H. \textit{et al}. Snapshots of laser wakefields. \textit{Nat. Phys.} \textbf{2}, 749-753 (2006).

\bibitem{Rice1994} Rice, A. \textit{et al}. Terahertz optical rectification from $<110>$ zinc-blende crystals. \textit{Appl.
Phys. Lett.} \textbf{64}, 1324-1326 (1994).

\bibitem{Chang1997} Chang, Z., Rundquist, A., Wang, H., Murnane, M. M. $\&$ Kapteyn, H. C. Generation of coherent soft X rays at
2.7 nm using high harmonics. \textit{Phys. Rev. Lett.} \textbf{79}, 2967-2970 (1997).

\bibitem{Popmin2012} Popmintchev, T. \textit{et al}. Bright coherent ultrahigh harmonics in the keV X-ray regime from
mid-infrared femtosecond lasers. \textit{Science} \textbf{336}, 1287-1291 (2012).


\bibitem{Thomson74}
Thomson, J. $\&$ Karush, J. I. Effects of finite-bandwidth driver on the parametric instability. \textit{Phys. Fluids} \textbf{17}, 1608-1613 (1974).

\bibitem{Zhao15}
Zhao, Y. \textit{et al}. Effects of large laser bandwidth on stimulated Raman scattering instability in underdense plasma.
\textit{Phys. Plasmas} \textbf{22}, 052119(1)-052119(7) (2015).

\bibitem{sideband}
\url{http://literature.agilent.com/litweb/pdf/5954-9130.pdf}

\bibitem{Sheng02}
Sheng, Z.-M., Mima, K., Sentoku, Y., Nishihara, K. $\&$ Zhang, J. Generation of high-amplitude plasma waves for particle acceleration by cross-modulated laser wake fields. \textit{Phys. Plasmas} \textbf{9}, 3147-3153 (2002).


\bibitem{Gupta99} Gupta, S. S.  $\&$ Kaw, P. K. Phase mixing of nonlinear plasma oscillations in an arbitrary mass ratio cold
plasma. \textit{Phys. Rev. Lett.} \textbf{82}, 1867-1870 (1999).

\bibitem{HXu06} Xu, H., Sheng, Z.-M. $\&$ Zhang, J. Phase mixing due to ion motion and relativistic effects in nonlinear
plasma oscillations. \textit{Phys. Scr.} \textbf{74}, 673¨C677 (2006).



\bibitem{mori1997} Mori, W. B. The physics of the nonlinear optics of plasmas at relativistic intensities for short-pulse lasers. \textit{IEEE J. Quan. Electr.} \textbf{33}, 1942-1953 (1997).

\bibitem{tzeng1999} Tzeng, K.-C., Mori, W. B. $\&$ Katsouleas T. Self-trapped electron acceleration from the nonlinear interplay between Raman forward scattering, self-focusing, and hosing. \textit{Phys. Plasmas}
\textbf{6}, 2105-2116 (1999).

\bibitem{wilks1989} Wilks, S. C., Dawson, J. M., Mori, W. B., Katsouleas, T. $\&$ Jones, M. E. Photon accelerator, \textit{Phys. Rev. Lett.} \textbf{62}, 2600-2603 (1989).

\bibitem{esarey1990} Esarey, E., Ting, A. $\&$ Sprangle, P.  Frequency shifts induced in laser pulses by plasma waves.
 \textit{Phys. Rev. A} \textbf{42}, 3526-3531 (1990).

\bibitem{sheng1993} Sheng, Z.-M., Ma, J.-X., Xu, Z.-Z. $\&$ Yu, W. Effect of an electron plasma wave on the propagation of an ultrashort laser pulse. \textit{J. Opt. Soc. Am. B} \textbf{10}, 122-129 (1993).

\bibitem{Ozimek87}
Ozimek, E. $\&$  S\c{e}k, A. Perception of amplitude and frequency modulated signals (mixed modulation). J. Acoust. Soc. Am. \textbf{82}, 1598-1603 (1987).

\bibitem{Kreh12}
Kreh, M. Bessel Functions \textit{G\"{o}tingen Summer School on Number Theory} 1-21 (2012).

\bibitem{Fonseca02}
Fonseca, R. A. \textit{et al}. OSIRIS, a three-dimensional
fully relativistic particle in cell code for modeling plasma based accelerators. \textit{Lect. Not. Comput. Sci.} \textbf{2331}, 342-351 (2002).

\end{thebibliography}
\end{document}